\documentclass[prd,twocolumn,showpacs,preprintnumbers,amsmath,amssymb,superscriptaddress,nofootinbib,english]{revtex4-1}

\usepackage{graphicx}
\usepackage{dcolumn}
\usepackage{bm}
\usepackage{epsfig}
\usepackage{graphicx}
\usepackage{hyperref}
\usepackage[usenames]{color}
\usepackage{url}
\usepackage{enumitem}
\usepackage{float}

\hypersetup{
    colorlinks=true,
    linkcolor=blue,
    citecolor=blue,
}

\newcommand{\remove}[1]{}

\def\ie{{\frenchspacing\it i.e.}}
\def\eg{{\frenchspacing\it e.g.}}

\def\be{\begin{equation}}
\def\ee{\end{equation}}
\def\ba{\begin{eqnarray}}
\def\ea{\end{eqnarray}}
\begin{document}
\title{A measurement of the Hubble constant using galaxy redshift surveys}

\author{Yuting Wang}
\email{ytwang@nao.cas.cn}
\affiliation{National Astronomy Observatories,
Chinese Academy of Science, Beijing, 100012, P.R.China}
\affiliation{Institute of Cosmology and Gravitation, University of Portsmouth,
Portsmouth, PO1 3FX, UK}

\author{Lixin Xu}
\email{lxxu@dlut.edu.cn}
\affiliation{Institute of Theoretical Physics,  School  of Physics \& Optoelectronic Technology, 
Dalian  University  of  Technology,  Dalian,  116024,  P.R.China}

\author{Gong-Bo Zhao}
\email{gbzhao@nao.cas.cn}
\affiliation{National Astronomy Observatories,
Chinese Academy of Science, Beijing, 100012, P.R.China}
\affiliation{Institute of Cosmology and Gravitation, University of Portsmouth,
Portsmouth, PO1 3FX, UK}

\begin{abstract}

We perform a measurement of the Hubble constant, $H_0$, using the latest baryonic acoustic oscillations (BAO) measurements from galaxy surveys of 6dFGS, SDSS DR7 Main Galaxy Sample, BOSS DR12 sample, and eBOSS DR14 quasar sample, in the framework of a flat $\Lambda$CDM model. Based on the Kullback-Leibler (KL) divergence, we examine the consistency of $H_0$ values derived from various data sets. We find that our measurement is consistent with that derived from Planck and with the local measurement of $H_0$ using the Cepheids and type Ia supernovae. We perform forecasts on $H_0$ from future BAO measurements, and find that the uncertainty of $H_0$ determined by future BAO data alone, including complete eBOSS, DESI and Euclid-like, is comparable with that from local measurements.

\end{abstract}


\maketitle

\section{Introduction}
Determining the Hubble constant, $H_0$, which is the present expansion rate of the Universe, with a high precision plays a crucial role in cosmology, and $H_0$ can be measured locally, or derived cosmologically through measurements of Cosmic Microwave Background (CMB) and Baryon Acoustic Oscillations (BAO) (see \cite{Freedman:2010xv, Freedman:2017yms} for a recent review on astronomical methods of $H_0$ measurements and its significance in cosmology).

Recently, a direct measurement of $H_0$ led by Riess \cite{Riess:2016jrr} (R16) using Cepheids and type Ia supernovae finds $H_0=73.24 \pm 1.74 \rm \,km\,s^{-1}\,Mpc^{-1}$, which is a 2.4\% measurement. On the other hand, a recent CMB measurement of $H_0$ using the Planck satellite (PLC15)  achieved a per cent level precision, namely, $H_0 =67.27\pm0.66 \rm \,km\,s^{-1}\,Mpc^{-1}$ \cite{Ade:2015xua}. Note that, unlike the local measurement, the CMB measurement of Hubble constant is model-dependent as a cosmological model, which is $\Lambda$CDM used for the measurement we quote here, is needed to convert the observed angular diameter distance at $z\sim1100$ and the sound horizon into a measurement of $H_0$. 

These two measurements are in apparent tension at more than $3\,\sigma$ level \cite{Riess:2016jrr}. The tension may imply that the $\Lambda$CDM used in the CMB analysis needs to be extended \cite{Wyman:2013lza, Wang:2015wga, Pourtsidou:2016ico, DiValentino:2016hlg, Zhao:2017cud, Sola:2017znb}, or that the measurements were contaminated by systematics to an unknown level. In this situation, additional independent measurements of $H_0$, \eg, using BAO distance measurements derived from galaxy surveys \footnote{There are other methods to determine the Hubble constant using galaxies. See \cite{Chen:2016uno} for an example.} \cite{Cheng:2014kja}, can provide critical information we need.

The BAO distance measurements using galaxy redshift surveys play a key role in probing the cosmic expansion history. The BAO characteristic scale can be measured in both radial and transverse directions of the line of sight to provide estimates of the Hubble parameter, $H(z)$, and angular diameter distance, $D_A(z)$, respectively at redshift $z$. Recently, the collaboration of Baryon Oscillation Spectroscopic Survey (BOSS), which is a part of Sloan Digital Sky Survey (SDSS)-III, performed BAO measurements in the redshift range of $0.2<z<0.75$ using the completed Data Release 12 (DR12) \cite{Acacia, BAOWang, BAOZhao}. The extended BOSS (eBOSS, part of SDSS-IV) detected a BAO signal at a 4\% precision at $z\sim1.5$ using the DR14 quasar sample \cite{Ata:2017dya}. These new BAO measurements can provide a $H_0$ measurement which is independent of CMB and local measurements, thus can be highly informative.  

In this paper, we determine the Hubble constant using the BOSS DR12 and eBOSS DR14 BAO measurements, combined with others available to date, and investigate the consistency of $H_0$ values derived from different data sets using the Kullback-Leibler (KL) divergence \cite{KL1951}. We also perform a forecast for future BAO data for a feasibility study.

This paper is organised as follows. In the next section, we present the method and data used in this work, followed by a section devoted to results. We present conclusion and discussions in Sec. IV. 

\section{Method and data} 

In the spatially flat $\Lambda$CDM model, the Hubble parameter is,
\ba
H(z) = H_0 \sqrt{\Omega_r(1+z)^4+\Omega_m(1+z)^3+\Omega_{\Lambda}},\,
\ea
where $\Omega_r+\Omega_m+\Omega_{\Lambda}=1$.The present energy density of radiation \footnote{Three species of massless neutrinos are included, and its energy density is given in terms of the photon density, $\rho_{\gamma}$, by $\rho_{\nu}=3(7/8)(4/11)^{4/3}\rho_{\gamma}$.} $\Omega_r=\Omega_m/(1+z_{\rm eq})$, with $z_{\rm eq}=2.5\times10^4\,\Omega_mh^2(T_{\rm CMB}/2.7 \,\rm K)^{-4}$, being the redshift of matter-radiation equality. We adopt $T_{\rm CMB}=2.7255 \,\rm K$. The angular diameter distance is, 
\ba
\label{da}
D_A(z) = \frac {1} {1+z} \int_0^z \frac {dz'} {H(z')}.\, 
\ea

\begin{figure*}[tbp]   
\includegraphics[scale=0.3]{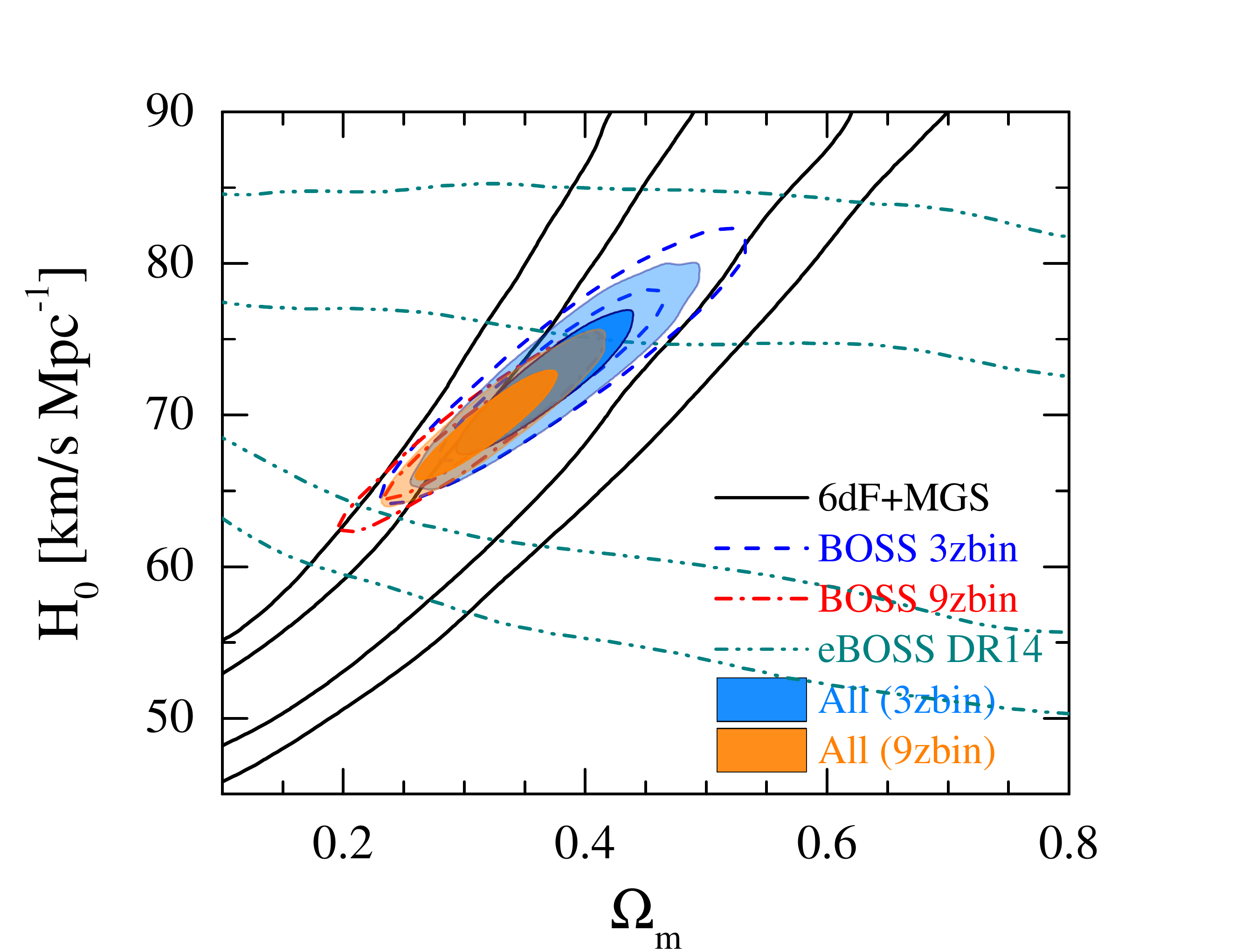}
\includegraphics[scale=0.3]{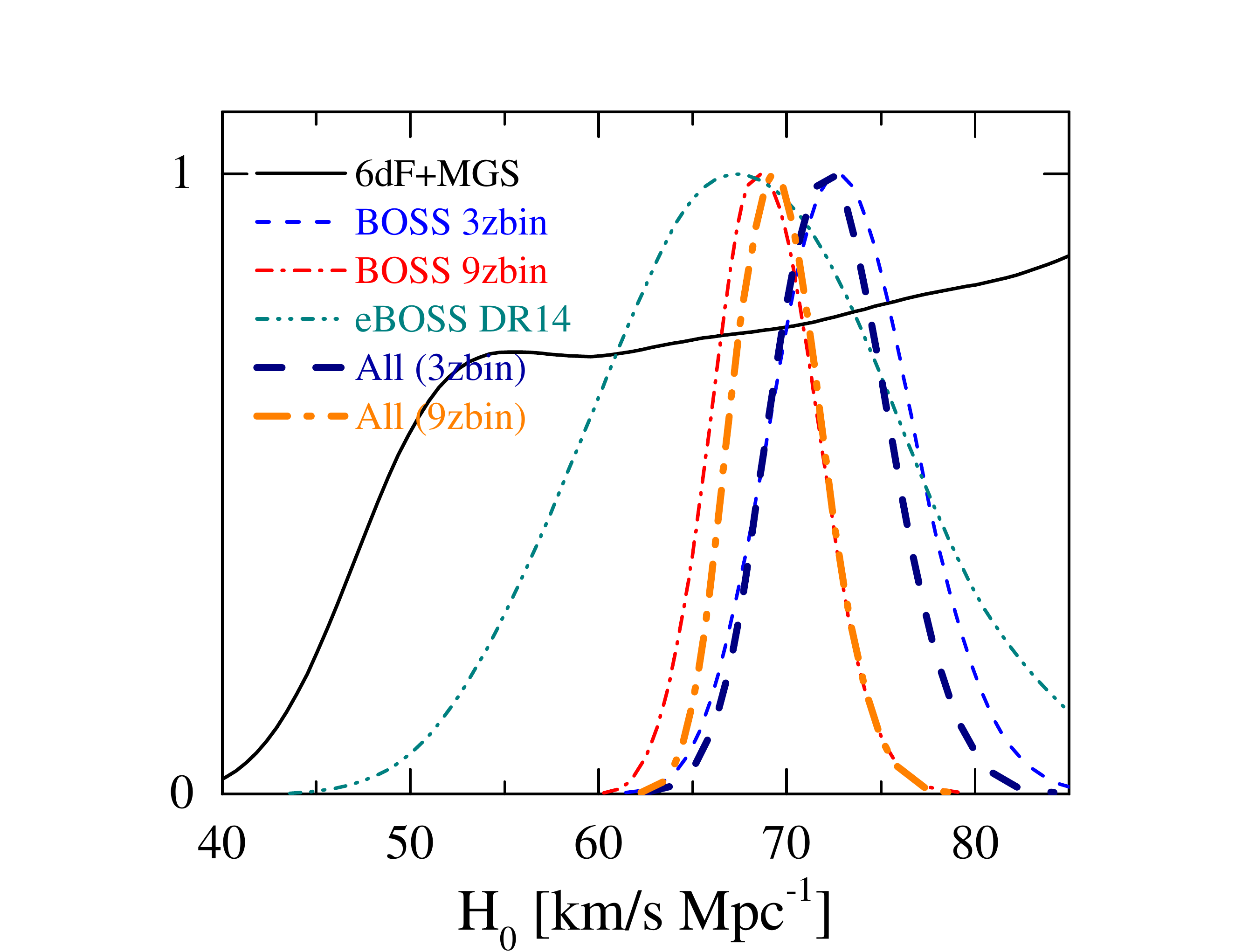}
\caption{Left panel: The 68 and 95\% confidence level (CL) contour plots of $\Omega_m$ and $H_0$ derived from various BAO measurements; Right panel: The probability distribution of $H_0$ derived from various BAO measurements.}\label{fig:LCDM_H0_Om}
\end{figure*} 

The sound horizon, $r_s$, at the redshift of the drag epoch, $z_d$, can be calculated as,
\ba
\label{eq:rs}
r_d \equiv r_s(z_d) =\int_0^{\frac 1 {1+z_d}} \frac {da'}  {a'^2H(a') \sqrt{3(1+\overline{R_b}a')}},\,
\ea
where $\overline{R_b}=3.15\times10^4\,\Omega_bh^2(T_{\rm CMB}/2.7 \,\rm K)^{-4}$. Note that $z_d$ is well approximated analytically \cite{Eisenstein:1997ik},
\ba
z_d = \frac {1291(\Omega_m h^2)^{0.251}}{1+0.659(\Omega_m h^2)^{0.828}}[1+b_1(\Omega_b h^2)^{b_2}],\,
\ea
where
\ba
b_1 &=& 0.313(\Omega_m h^2)^{-0.419}[1+0.607(\Omega_m h^2)^{0.674}],\, \\
b_2 &=& 0.238(\Omega_m h^2)^{0.223}.\, 
\ea
We use a fixed value of the baryon density $\Omega_b h^2=0.02225$ from the Planck result \cite{Ade:2015xua} \footnote{We have tested to marginalise over $\Omega_b h^2$ with a Gaussian prior derived from the Planck measurement, and find that the result is largely unchanged.}. The baryon density can also be accurately determined in a CMB-independent way, \eg, using the primordial deuterium abundance in Big Bang Nucleosynthesis (BBN) theory \cite{Riemer-Sorensen:2017vxj}. 

Note that the quantities $H(z)r_d$ and $D_A(z)/r_d$ can be estimated from anisotropic BAO measurements, while the quantity, 
\ba
D_V(z)/r_d \equiv \left [(1+z)^2D_A^2(z)\frac {z} {H(z)}\right]^{\frac 1 3} /r_d, 
\ea
is determined by isotropic BAO measurements. 

As shown above, the BAO distance measurements, $H(z)r_d, D_A(z)/r_d$ or $D_V(z)/r_d$ are two-variable functions of $\Omega_m$ and $H_0$ (once $\Omega_b h^2$ is known) in a flat $\Lambda$CDM cosmology, therefore the Hubble constant can be in principle determined from the BAO distances with $\Omega_m$ marginalised over. 

In what follows, we use isotropic or anisotropic BAO distance measurements to determine the Hubble constant with $\Omega_m$ marginalised over, \ie, our parameter space is simply (assuming a flatness of the Universe),
\ba
P\equiv \{\Omega_m, H_0\}\ 
\ea
The sound horizon at the drag redshift $r_d$ is calculated using Eq. \ref{eq:rs} \footnote{Except for 6dF, the value of $r_d$ is rescaled by a factor $~r_d/\tilde{r}_d$, where $\tilde{r}_d$ value is calculated from CAMB \cite{CAMB} in the same fiducial cosmology \cite{Bennett:2014tka}.}. We perform a  Monte Carlo Markov Chain (MCMC) global fitting for parameter estimation using a modified version of CosmoMC \cite{Lewis:2002ah} \footnote{\url{http://cosmologist.info/cosmomc/}}.

The BAO datasets used in this work include,

\begin{itemize}
\item The isotropic BAO measurements using the 6dFRS (6dF) \cite{6df} and SDSS main galaxy sample (MGS) \cite{MGS} at effective redshifts $z_{\rm eff}=0.106$ and $z_{\rm eff}=0.15$ respectively;
\item  The BOSS DR12 anisotropic BAO measurements at three effective redshifts (BOSS 3$z$bin) in \cite{Acacia} or at nine effective redshifts (BOSS 9$z$bin) in \cite{BAOWang, BAOZhao};
\item  The eBOSS DR14 isotropic BAO measurement at $z_{\rm eff}=1.52$ \cite{Ata:2017dya};
\item  A combination of 6dF + MGS + BOSS 3$z$bin + eBOSS DR14 (All 3$z$bin), or a combination of 6dF + MGS + BOSS 9$z$bin + eBOSS DR14 (All 9$z$bin).
\end{itemize}

To check the consistency of $H_0$ values determined from different data sets within the $\Lambda$CDM model, we compute the tension $T$ based on the KL divergence \cite{Kunz:2006mc, Paykari:2012ne, Amara:2013swa, Seehars:2014ora, Verde:2014qea, Grandis:2015qaa, Seehars:2015qza, Raveri:2016xof}, which quantifies the distance between two probability density functions (PDFs), $p_1$ and $p_2$. If both $p_1$ and $p_2$ are assumed to be Gaussian, the relative entropy in bits between the two PDFs can be evaluated as,

\begin{equation}
	\begin{aligned}
		D(p_2||p_1) &=\frac{1}{2 \log 2}  \left[ \text{Tr}\left( \mathcal{C}_1^{-1}  \mathcal{C}_2\right) - d - \log \frac {\det  \mathcal{C}_2} {\det  \mathcal{C}_1} \right.\\
		&\qquad + \left.(\theta_2 - \theta_1)^T\mathcal{C}_1^{-1}(\theta_2 - \theta_1)\right], \\
	\end{aligned}
\end{equation}
where $\theta_i$ is the best-fit parameter vector, $\mathcal{C}_i$ is the corresponding covariance matrix, and $d$ denoted the dimensions of the parameter space (\eg, $d=2$ in our case where both $H_0$ and $\Omega_m$ are free parameters). If data are assumed to be more informative than the priors, one can compute the expected relative entropy, $\langle D \rangle$, with its standard deviation, $\Sigma$, via,
\begin{eqnarray}
	\langle D \rangle & \simeq& \frac{1}{\log 2} \left[ \text{Tr} (\mathcal{C}_2  \mathcal{C}_1^{-1}) - \frac{1}{2} \log \frac {\det  \mathcal{C}_2} {\det  \mathcal{C}_1} \right],\\
	\Sigma(D) &\simeq& \frac{1}{\sqrt{2}\log 2} \sqrt{\text{Tr} \left(\mathcal{C}_1^{-1} \mathcal{C}_2 + \mathbb{I} \right)^2 }, \\
	S &\equiv&  D(p_2||p_1)- \langle D \rangle, 
\end{eqnarray}
where the Surprise, $S$, is defined as the difference between the relative entropy and its expectation value. The tension, $T$, is defined as the signal-to-noise ratio of the Surprise, \ie \,
\begin{eqnarray} \label{eq:tension}
T  \equiv S/ \Sigma. 
\end{eqnarray}
If $T\lesssim1$, then $p_1$ and $p_2$ are consistent with each other, while otherwise the two PDFs are in tension \cite{Seehars:2015qza}.

We also perform forecasts on the uncertainty of $H_0$ using ongoing and upcoming redshift surveys, including eBOSS \footnote{We use ``eBOSS'' throughout for the complete 5-year eBOSS sample, while use ``eBOSS DR14'' to denote the eBOSS DR14 quasar sample. More information of the eBOSS survey is available here: \url{http://www.sdss.org/surveys/eboss/}} \cite{Dawson:2015wdb, Zhao:2015gua}, Dark Energy Spectroscopic Instrument (DESI) \footnote{\url{http://desi.lbl.gov/}} \cite{Aghamousa:2016zmz, Aghamousa:2016sne}, and ESA's Euclid satellite \footnote{\url{http://www.euclid-ec.org/}}  \cite{Euclid}. We use a flat, $\Lambda$CDM cosmology derived from the Planck mission as our fiducial model \cite{Ade:2015xua}, take the forecasted BAO data for galaxy surveys of a complete eBOSS from \cite{Zhao:2015gua} (\ie\, the BAO result from eBOSS Luminous Red Galaxies, High Density Emission Line Galaxies and Clustering quasars in Table 4 from \cite{Zhao:2015gua}), DESI (\ie\, the BAO result from DESI Luminous Red Galaxies, Emission Line Galaxies and Clustering quasars in Table 2.3 from \cite{Aghamousa:2016zmz} and DESI Bright Galaxies in Table 2.5 from \cite{Aghamousa:2016zmz}) and Euclid-like \cite{Font-Ribera:2013rwa} (\ie\, Table VI in \cite{Font-Ribera:2013rwa}) respectively, and perform parameter estimation using the MCMC method, in the same way as we did for current datasets.

\section{Results} 
\begin{table}
\begin{center} 
\begin{tabular}{ccc}
\hline  \hline   
 Dataset        &     $H_0 [\rm km/s\, Mpc^{-1}]$ & precision\\ \hline
 All 3$z$bin  &      $71.75\pm3.05$  & $4.25$ \% \\ 
 All 9$z$bin  &      $69.13\pm2.34$  & $3.38$ \% \\ 
 R16		   &	    $73.24\pm1.74$  & $2.38$ \% \\
 PLC15        &       $67.27\pm0.66$  & $0.98$ \% \\
 \hline
 eBOSS       &       $67.27\pm1.55$  & $2.30$ \% \\
 DESI          &       $67.27\pm0.33$  & $0.49$ \% \\
 Euclid-like         &       $67.27\pm0.21$  & $0.31$ \% \\
\hline  \hline                       
\end{tabular}
\end{center}
\caption{The mean and 68\% CL constraint on $H_0$ using various datasets. The upper part of the table (above the horizontal line) is for current datasets, while the lower part shows the forecast result based on a fiducial model derived from PLC15.}
\label{tab:result}
\end{table}

We present the joint constraint on $H_0$ and $\Omega_m$, and the posterior probability distribution of $H_0$ from various BAO datasets, including the latest eBOSS DR14 quasar sample, in Figure \ref{fig:LCDM_H0_Om}. As shown, the contours derived from different datasets show different degeneracy between $H_0$ and $\Omega_m$. This is expected as the degeneracy is largely determined by the effective redshift at which the BAO measurement is performed. Hence having tomographic BAO measurements at a large number of redshifts helps to break the degeneracy. This can be seen by comparing the ``All 3$z$bin" to ``All 9$z$bin" results. The only difference in these two datasets is that the BOSS DR12 galaxies were subdivided into more redshift slices in the ``9$z$bin" sample to gain more light-cone information. As shown in the upper part of Table \ref{tab:result}, the improvement on the uncertainty of $H_0$ is significant, namely, the error of $H_0$ reduces from $3.05$ to $2.34$ $\rm \,km\,s^{-1}\,Mpc^{-1}$, which is a 23\% improvement.

\begin{table}
\begin{center} 
\begin{tabular}{cccccc}
\hline  \hline   
 Dataset                                  &     $D$ &$\langle D \rangle$& $S$ & $T$ & $\sigma_{T}$ \\  \hline
2D: $\{\Omega_m, H_0 \}$      &&&&& \\
 All 3$z$bin $\to$  All 9$z$bin  &      $0.87$ & $2.46$& $-1.59$& $-0.71$& $0.14$\\ 
 All 3$z$bin $\to$  PLC15        &       $7.91$ & $7.87$& $0.04$& $0.02$ & $1.12$\\ 
 All 9$z$bin $\to$  PLC15	      &	       $8.58$ &  $8.31$& $0.27$&$0.09$ &$1.54$\\  \hline
 1D: $\{ H_0 \}$      &&&&& \\
 All 3$z$bin $\leftrightarrow$  All 9$z$bin  &      $0.62$ & $1.23$& $-0.61$& $-0.38$& $0.19$\\ 
 All 3$z$bin $\leftrightarrow$  PLC15        &       $3.08$ & $2.28$& $0.80$& $0.75$ &$1.74$\\ 
 All 9$z$bin $\leftrightarrow$  PLC15	       &	$1.62$ &  $1.94$& $-0.32$&$-0.29$ &$0.94$\\  
 All 3$z$bin $\leftrightarrow$  R16            &       $0.50$ &  $1.28$& $-0.78$&$-0.58$& $0.26$\\
 All 9$z$bin $\leftrightarrow$  R16             &      $2.33$ &  $1.23$& $1.10$&$0.70$& $0.42$\\ 
 PLC15	 $\leftrightarrow$  R16             &        $61.91$ &  $8.63$& $53.29$&$6.57$&$2.73$ \\ 
\hline  \hline                       
\end{tabular}
\end{center}
\caption{Top table: The KL divergence between the PDFs for $\Omega_m$ and $H_0$ using BAO data and PLC15. Bottom table: The KL divergence between the PDFs for $H_0$ with $\Omega_m$ marginalised over from various datasets. The tension, $T\lesssim1$, illustrates the relevant pairs of datasets are consistent with each other.}
\label{tab:KL_2D}
\end{table}

 \begin{figure}[tbp]   
\includegraphics[scale=0.27]{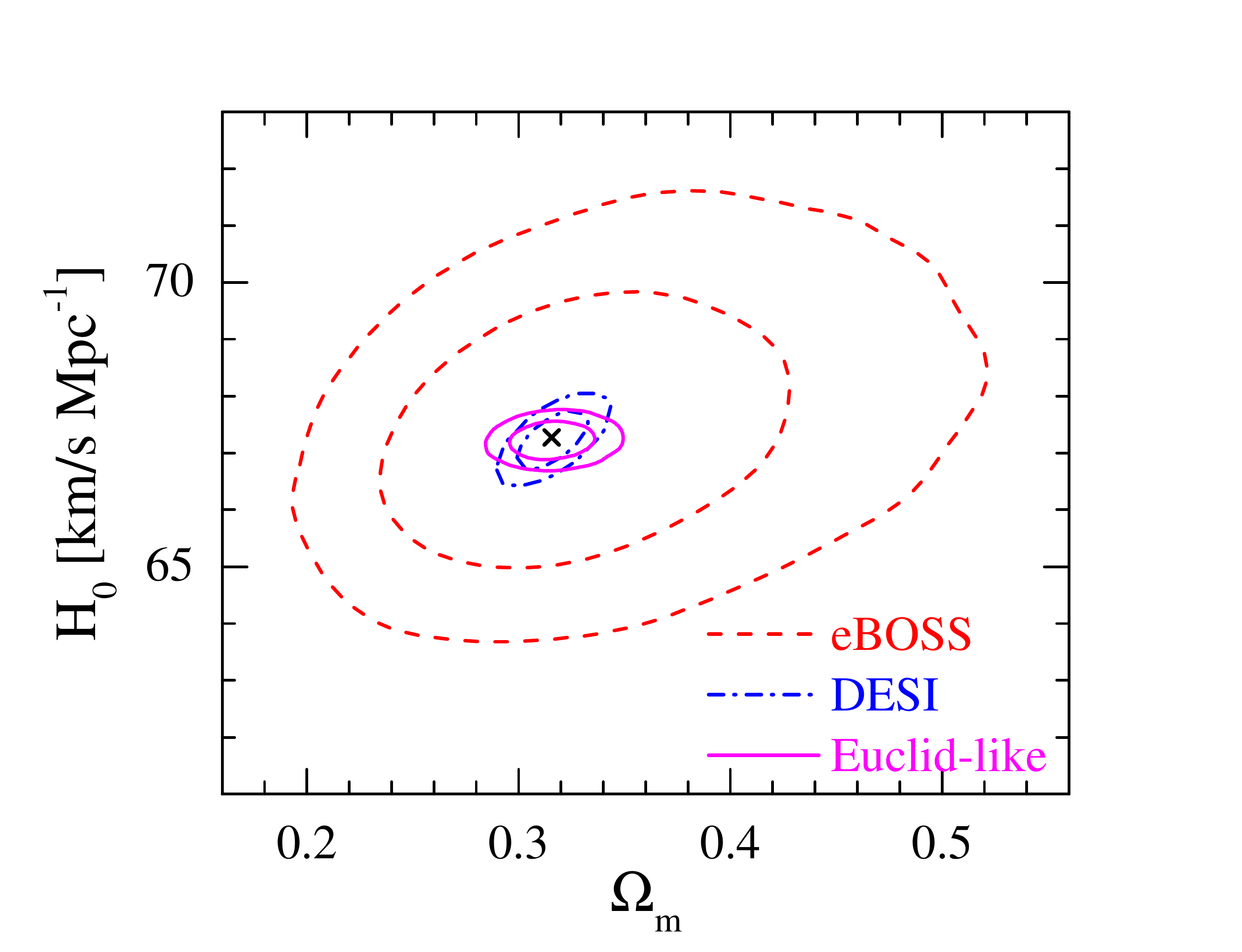}
\caption{The 68 and 95\% CL contour plots of parameters $\Omega_m$ and $H_0$ in a flat $\Lambda$CDM model, derived from the complete eBOSS (red dashed), DESI (blue dash-dotted) and Euclid-like (magenta solid). The black cross corresponds to the fiducial model.}\label{fig:Future_9z}
\end{figure} 

We quantify the (in)consistency among the derived $\Omega_m$ and $H_0$ from BAO data and PLC15, using the quantity defined in Eq (\ref{eq:tension}). We also calculate the KL divergence between the PDFs for $H_0$ with $\Omega_m$ marginalised over from various datasets, including those from PLC15 and R16. The result is presented in Table \ref{tab:KL_2D}, including the relative entropy, $D$, its expected value, $\langle D \rangle$, the Surprise, $S$ in bits, and the tension, $T$ with $1\,\sigma$ error. As shown, except for the PLC15 and R16 pair, where $T$ is larger than $1$ at about $2\,\sigma$ level, all others are consistent with each other (the tension $T$ are all less than unity).

Given that the best measurement of $H_0$ to date using BAO alone (\ie, the ``All 9$z$bin" result) has a worse precision than R16 or PLC15, we investigate the constraining capability of future BAO surveys, including the complete eBOSS, DESI and Euclid-like, on $H_0$. The joint constraint on $H_0$ and $\Omega_m$, and the marginalised constraint on $H_0$, from these surveys are shown in Figure \ref{fig:Future_9z}, and in the lower part of Table \ref{tab:result}, respectively. As shown, future galaxy surveys, especially for DESI or Euclid-like alone, is able to provide a better constraint on $H_0$ than the current CMB constraint, which is promising. 

\section{Conclusion and Discussions}

In this paper, we determine the Hubble constant using BAO measurements from galaxy redshift surveys in a flat $\Lambda$CDM cosmology. A combination of recent BAO measurements from 6dF, MGS, BOSS DR12 (with $9$ redshift slices) and eBOSS DR14 quasar sample yields a measurement of Hubble constant, namely, $H_0 = 69.13 \pm 2.34 \rm \,km\,s^{-1}\,Mpc^{-1}$, which is a 3.4\% measurement. Given level of the uncertainty, this measurement is consistent with both R16 and PLC15, which are in tension between themselves. 

Based on a forecast, we find that future galaxy surveys including DESI and Euclid-like, will be able to provide competitive constraints on $H_0$, compared with current local or CMB measurements.

\begin{acknowledgments}
We thank Will Percival for discussions and comments. We also thank Chao Liu for discussions. YW is supported by the NSFC Grant No. 11403034, and by the Young Researcher Grant of National Astronomical Observatories, Chinese Academy of Sciences. LX is supported by the NSFC Grant No. 11275035, Grant No.11675032, and by ``the Fundamental Research Funds for the Central Universities" under Grant No. DUT16LK31. GBZ is supported by NSFC Grant No. 11673025, and by a Royal Society-Newton Advanced Fellowship.  

This research used resources of the SCIAMA cluster supported by University of Portsmouth, and the ZEN cluster supported by NAOC.
\end{acknowledgments}

\end{document}